\documentclass{iau} 
\usepackage{graphicx}
\usepackage{natbib}

\title[Physical Conditions of Planet Formation] 
{Tracing The Physical Conditions of Planet Formation with Molecular Excitation}

\author[Richard Teague]   
{Richard Teague$^1$}

\affiliation{$^1$Department of Astronomy \\
             University of Michigan \\
             1085 S. University Ave. \\
             Ann Arbor, MI 48109, USA \\
             email: {\tt rteague@umich.edu}}

\pubyear{2019}
\volume{S350}  
\jname{Laboratory Astrophysics: From Observations to Interpretation}
\editors{A.C. Editor, B.D. Editor \& C.E. Editor, eds.}
\begin{document}

\maketitle

\begin{abstract}
Understanding the physical structure of the planet formation environment, the protoplanetary disk, is essential for the interpretation of high resolution observations of the dust and future observations of the magnetic field structure. Observations of multiple transitions of molecular species offers a unique view of the underlying physical structure through excitation analyses. Here we describe a new method to extract high-resolution spectra from low-resolution observations, then provide two case studies of how molecular excitation analyses were used to constrain the physical structure in TW~Hya, the closest protoplanetary disk to Earth.
\end{abstract}

\firstsection 
\section{Introduction}

Protoplanetary disks are the birthplace of planets. With a significant amount of substructure observed both in the dust and the gas \citep[e.g.][]{Andrews18}, it is highly likely that with the Atacama Large (sub-)Millimeter Array (ALMA) we are witnessing first-hand the assembly of planetary systems. Relating these observed structures to the planet formation process, however, requires intimate knowledge of the underlying physical conditions, namely the gas temperature and densities, both of which strongly influence the pace of planet formation and the subsequent interaction of the protoplanets with the disk.

Observations of multiple transitions of common molecular tracers offer an opportunity to measure these physical conditions. Although excitation analyses \citep[such as a population diagram analyses;][]{Goldsmith99} are commonplace in studies of the ISM and earlier stages of star formation, application to protoplanetary disks is hampered by the low intensities of the molecular lines and the high spatial resolution required to spatially resolve the disk. Nonetheless, the significant improvement in sensitivity afforded by ALMA has allowed for multiple such analyses, such as \citet{Schwarz16}, \citet{Bergner18}, \citet{Loomis18b} and \citep{Teague18a}, which are demonstrating the power of such methods.

\section{Extracting Spectroscopy-Ready Spectra}
\label{sec:method}

Although ALMA has undoubtedly revolutionised the study of planet formation and protoplanetary disks, the intrinsically weak emission of most molecular species due to their low column-densities limits the type of spectral analyses possible. The development of new techniques to detect weak lines, such as match filtering \citep{Loomis18a}, are essential to maximise the information we are able to extract from observations. As protoplanetary disks are observed to be predominantly azimuthally symmetric \citep[see the DSHARP survey, for example;][]{Andrews18}, and have a well defined dynamical structure, it is possible to leverage this to improve the quality of spectra extracted from the observations.

\begin{figure}
    \centering
    \includegraphics[width=\textwidth]{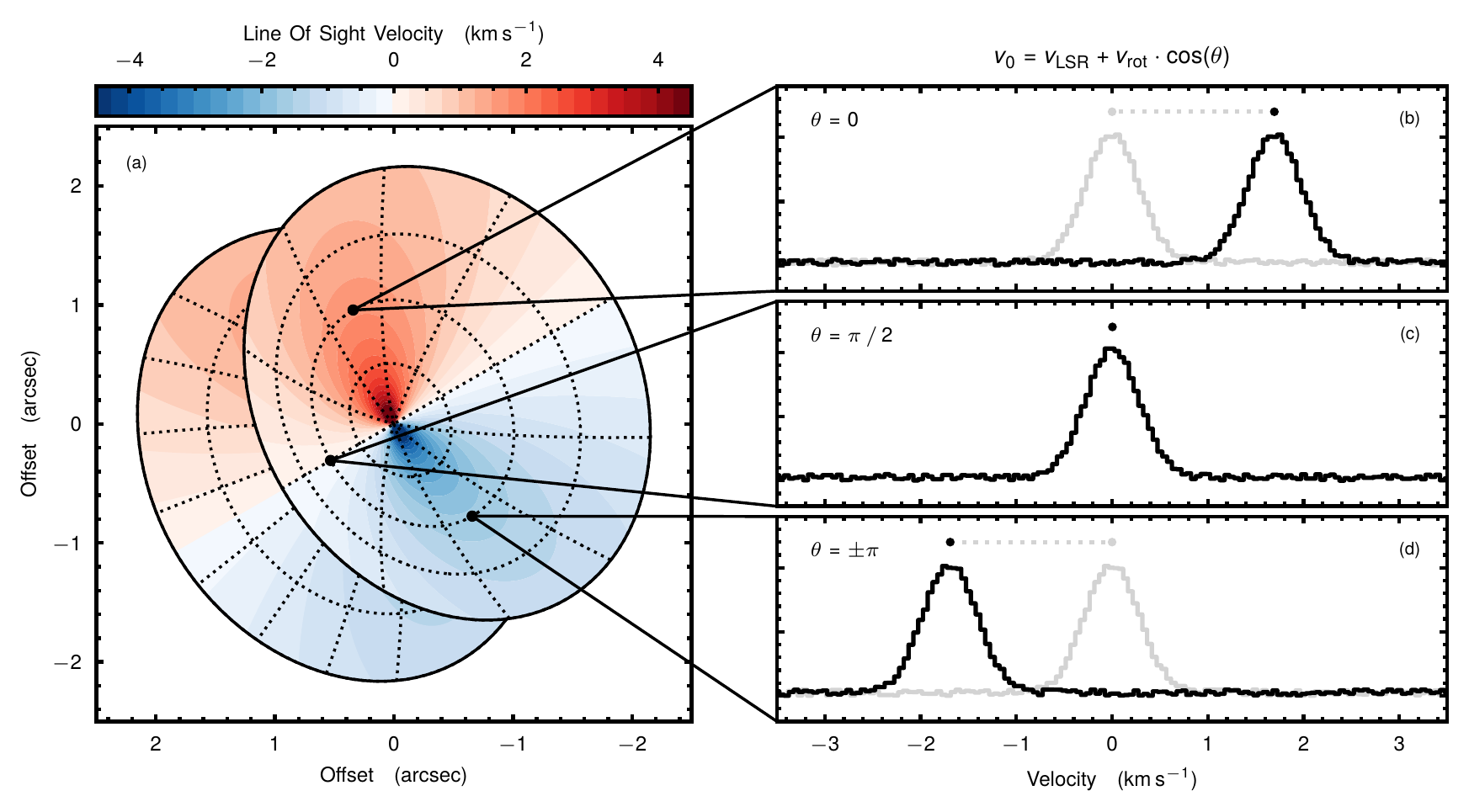}
    \caption{A cartoon demonstrating how spectra are Doppler shifted due to the rotation of the disk. The left panel shows a rotation map of a model disk, showing both the near and far sides of a flared emission surface. The panels on the right show in black the spectra extracted at three locations which are able to be shifted back to the line center given Eqn.~\ref{eq:vlos}. Taken from \citet{Teague18a}.}
    \label{fig:first_moment_example}
\end{figure}

Due to the rotation of the disk, emission lines will be Doppler shifted relative to the systemic velocity by,

\begin{equation}
    v_{\rm los}(r,\,\phi) = v_{\phi}(r) \cdot \cos \phi \cdot \sin i,
    \label{eq:vlos}
\end{equation}
\\
\noindent where $v_{\phi}$ is dominated by Keplerian rotation, $v_{\rm kep}$, $\phi$ is the polar angle at the disk midplane relative to the redshifted major-axis of the disk, and $i$ is the inclination of the disk. As the geometrical properties of the disk are readily measured from continuum observations, it is possible to infer $v_{\rm los}$ for each pixel and thus shift spatially resolved spectra back to the systemic velocity, as shown in Fig.~\ref{fig:first_moment_example}. With all the spectra aligned, it is then possible to stack spectra over a given region, for example around an annulus of constant radius, and thus improve the signal-to-noise ratio of the spectrum. \citet{Yen16} applied this technique to the disk of HD~163296, extracting detections of H$_2$CO which were not found with traditional imaging techniques, and significantly improving the significance of detection for both DCO$^+$ and N$_2$D$^+$.

An additional advantage of this process is that the spectra can be super-sampled when stacking the multiple components. As $v_{\rm los}$ is not discretized, unlike the spectral resolution of the telescope, the shifted spectra will end up sampling the intrinsic line profile at a spectral resolution roughly a factor $\sqrt{N}$ better, where $N$ is the number of independent spectra used in the stacking \citep{Teague18b}. It is important to note that this super-sampled spectrum will still contain any systemtic effects found in the intrinsic spectrum, for example any broadening due to the spectral response function \citep[e.g.][]{Koch18}, and will therefore only be able to recover the true intrinsic spectrum if the original data was taken at a spectral resolution sufficient to Nyquist-sample the line profile. The Python package \texttt{GoFish} \citep{Teague19} provides the necessary functions to split a disk into annuli and then align the spectra given a $v_{\phi}$ profile.

\section{Application to TW~Hya: Two Case Studies}
\label{sec:TWHya}

TW~Hya is the closest protoplanetary disk to Earth at a distance of $60.1 \pm 0.1$~pc \citep{BailerJones18}, and is therefore an object of intense study. With an inclination of $i \sim 7^{\circ}$, the geometry is exceptionally favourable for excitation analyses as confusion from radial gradients in temperature and densities are minimised. In the following section, we present two case studies of how physical properties of the planet forming disk were extracted through molecular excitation analyses.

\subsection{Surface density perturbations traced with CS}
\label{sec:TWHya:CS}

TW~Hya is known to host a significant amount of gap and ring substructures, both in the mm~continuum \citep{Andrews16} and in the NIR scattered light \citep{vanBoekel17}, suggestive of embedded protoplanets. However, as the distribution of dust is strongly dependent on the gas pressure gradient, it is essential to constrain the true gas density profile by using molecular emission to more accurately constrain the depth of the gap and thus the mass of any potential planet.

\citet{Teague17} showed that the CS $J = 5-4$ transition exhibits a gap in its radial intensity profile, consistent in location with the $\approx 95$~au gap observed in the scattered NIR light. Thermo-chemical modelling of the line emission suggested that a surface density depletion of 55\% was the most likely scenario, requiring a planet ranging between 12 -- $38~M_{\rm Earth}$. However, with only a single transition, breaking the degeneracy between temperature and column density was impossible, meaning that a chemical or excitation scenario could not be ruled out.

With the addition of the CS $J = 3-2$ and $J = 7-6$ transitions, \citet{Teague18a} were able to perform an excitation analysis assuming local thermodynamic equilibrium (LTE) to place limits on $T_{\rm ex}$, finding temperature of $\approx 40$~K in the inner disk, dropping to $\lesssim 20$~K at the disk edge of 200~au. Drops in the column density of CS were found to coincide with the previously claimed drop in total gas surface density. Using the resampling technique described in \S\ref{sec:method}, high resolution spectra were extracted for annuli spanning the full radius of the disk, each with a width of $0.15^{\prime\prime}$ ($\approx 9$~au). With these spectra, a full non-LTE excitation analysis was performed using \texttt{RADEX} \citep{vanderTak07}, the collisional rates from \citet{Lique06} and assuming a thermal H$_2$ ortho-to-para ratio. Radial profiles of $T_{\rm ex}$ and $N({\rm CS})$ consistent with the LTE analysis were found over most of the disk, demonstrating that these transitions were thermalised and that CS emission arises from a relatively dense region of the disk. 

Interestingly, it was found that in the outer 20~au of the disk, the $J = 7-6$ transition appeared to no longer be thermalised, requiring a collider density of $n({\rm H_2}) \sim 10^6~{\rm cm^{-3}}$. It was possible to extrapolate this to a gas surface density under the assumption that CS arises from a region close to the disk midplane \citep[as suggested by observations of the edge-on disk, the Flying Saucer;][]{Dutrey17}, finding a minimum $\Sigma_{\rm gas} \gtrsim 10^{-2}~{\rm g\,cm^{-2}}$ at 200~au. This value is almost two orders of magnitude larger than that from \citet{Bergin13}, who were modelling HD emission from the inner disk, but consistent with the models of \citet{vanBoekel17} who were able to reproduce scattered light out to the disk edge. This suggests that disks may contain more mass in their outer regions than previously thought, supporting a large reservoir of planet-building material at large radii.

\subsection{Where does CN emission arise from?}
\label{sec:TWHya:CN}

Constraining the location of the CN emission is essential for up-coming observations of polarized CN emission which aim to trace the projected magnetic field strength and morphology. The structure of the magnetic field is believed to change as a function of height through the disk as the initially poloidal fields are dragged into a toroidal morphology at the disk midplane due to the large gas densities. Therefore, knowing the region traced by the observed emission is fundamental in interpreting such observations of polarized emission.

It is currently under debate where CN arises in a protoplanetary disk. \citet{HilyBlant17} present an LTE analysis of the $N = 3-2$ transition in TW~Hya finding a low $T_{\rm ex}$ ranging between 17 -- 27~K, consistent with a similar LTE analysis of the $N = 2-1$ transition in \citet{Teague16} \citep[and those from other disks;][]{Chapillon12}, suggesting an emission region closer to the disk midplane where $z/r \lesssim 0.25$. Conversely, \citet{Cazzoletti18} found that the emission morphology, a single bright ring, was best reproduced when CN was formed via vibrationally excited H$_2$, requiring high levels of FUV radiation and therefore favouring a higher emission surface of $z/r \sim 0.4$.

\begin{figure}[t]
    \centering
    \includegraphics[width=\textwidth]{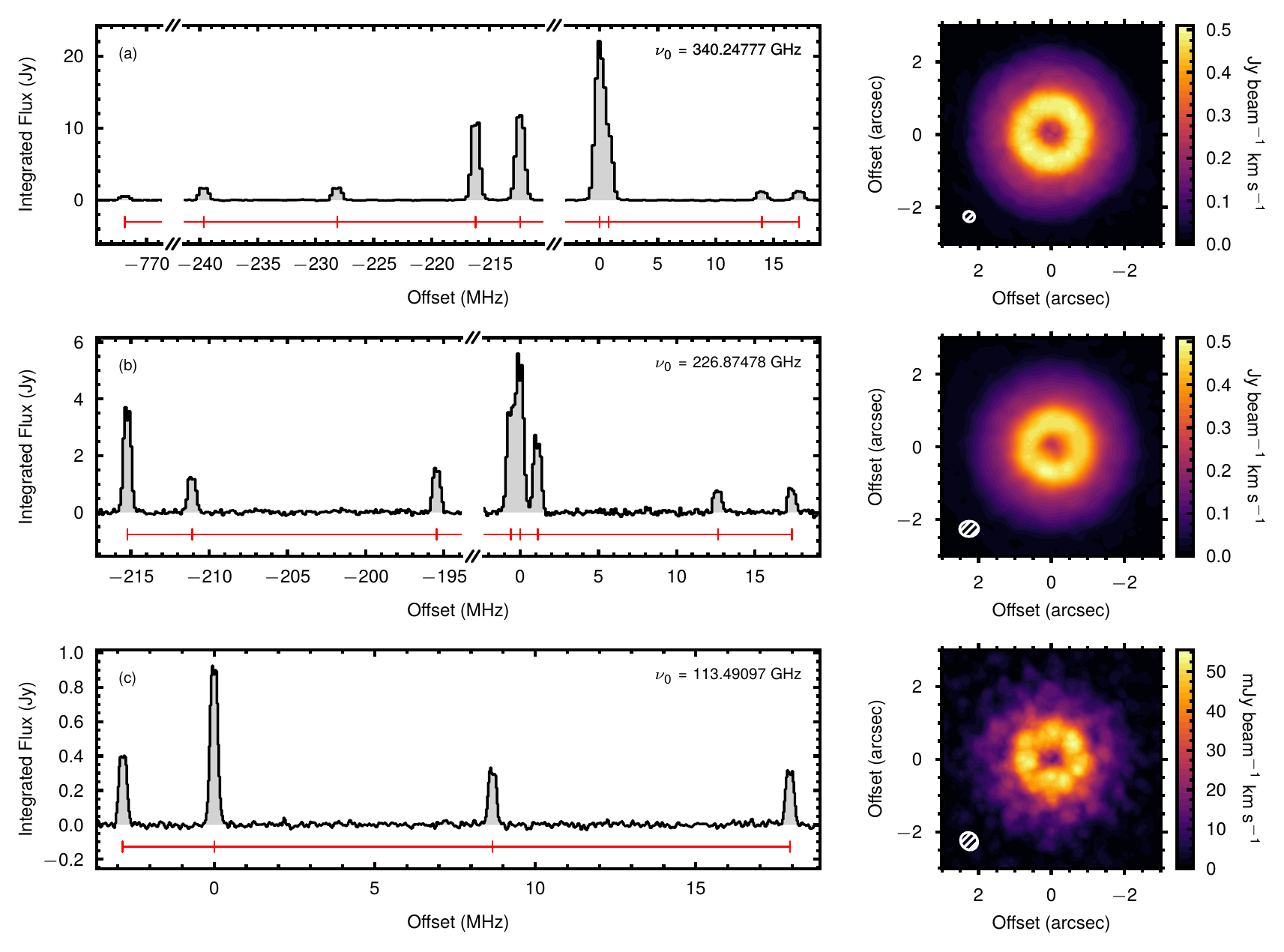}
    \caption{CN emission from TW~Hya. The left panels show the integrated flux, while the right panels show the zeroth moment maps with the synthesized beam in the bottom left of each panel. The rows show the $N = 3-2$, $N = 2-1$ and $N = 1-0$ transitions, top to bottom.}
    \label{fig:CN_observations}
\end{figure}

Archival ALMA data now exists for CN with data spanning the $N = 1-0$, $N = 2-1$ and $N = 3-2$ transitions allowing for a full excitation analysis in order to place tighter limits on the emission region. Fig.~\ref{fig:CN_observations} shows the three transitions and the mutliple spatially and spectrally resolved hyerpfine transitions detected (Loomis et al., in prep.). All three transitions display a similar emission morphology, a ring centered at 55~au, suggesting that this structure is due to a chemical effect rather than an excitation effect.

Applying the LTE excitation analysis to the super-sampled spectra using the level structure from \citet{Kalugina15}, it was found that the weaker satellite hyperfine components were unable to be reproduced, most significantly for the $N = 3-2$ transition, and to a lesser extent for the two lower frequency transitions. This can be attributed to the fact that in all previous analyses, the Rayleigh-Jeans approximation was assumed to convert between flux density units and brightness temperature. However, at these frequencies, $h\nu \lesssim kT$, limiting the accuracy of this transformation. Using the full Planck law, it was found that large optical depths, $\tau \gg 1$, were needed to explain the integrated intensities, producing saturated line profile inconsistent with the high spectral-resolution data.

Furthermore, the measured linewidths for all three transitions suggest $T_{\rm kin} > T_{\rm ex}$, even after correcting for systematic effects \citep{Teague16}. Previous estimates of the non-thermal broadening in TW~Hya show that $v_{\rm turb} \, / \, c_s < 10^{-1}$, where $c_s$ is the local gas sound speed \citep{Teague16, Flaherty18}, insufficient to explain the discrepancy between these two temperatures. Combined with the inability of the LTE model to reproduce the emission, this strongly suggests that these transitions are sub-thermally excited, and thus require lower densities of $n({\rm H_2}) \lesssim 10^6~{\rm cm^{-3}}$ \citep{Shirley15}, suggesting an emission layer of $z \, / \, r \gtrsim 0.3$ \citep{HilyBlant17}. A non-LTE excitation analysis would provide tighter constraints on the local H$_2$ density and will be presented in future work (Teague et al., in prep.)

\section{Summary}
\label{sec:summary}

We have demonstrated a new technique to extract high-spectral resolution data from lower spectral resolution observations by exploiting the azimuthal symmetry of the protoplanetary disk and the dynamics of the gas known to be dominated by Keplerian rotation \citep{Teague18b}. We have also shown two case studies of how observations of multiple transitions of molecular line emission allow us to begin to uncover the underlying physical structure of the planet formation environment. Lines of CS were able to confirm the existence of a surface density depletion in the disk of TW~Hya, likely opened by an unseen embedded planet \citep{Teague18a}. In addition, with high spectral resolution observations of CN we were able to demonstrate that the transitions were sub-thermally excited, and thus must originate from higher regions in the disk, unlike previous estimations (Teague et al., in prep.).

\end{document}